\documentstyle[psfig,draft]{mn}

\title[Evolution of 1612\,MHz Maser Emission in Expanding Circumstellar Shells]
      {Evolution of 1612\,MHz Maser Emission in Expanding Circumstellar Shells}

\author[M.\,D.\ Gray, D.\,A.\ Howe et al.]
       {M.\,D.\ Gray$^{1}$, D.\,A.\ Howe$^{1}$ \& B.\,M.\ Lewis$^{2}$\\ 
        $^{1}$ Department of Physics, University of Manchester, Sackville St.
	       Building, PO Box 88, Manchester, M60 1QD, UK\\
        $^{2}$ Arecibo Observatory, HC3, Box 53995, Arecibo, PR 00612, USA}

\date{Accepted ... .
      Received ... ;
      in original form ...}

\pagerange{000--000}

\begin{document}

\maketitle

\begin{abstract}
Observations show that 1612\,MHz masers of OH/IR
stars can fade on a timescale of a decade. This fading
is probably associated with the switch from rapid mass loss, 
which is ultimately linked
with an internal He-shell flash, to the much slower mass loss supported by
more quiescent conditions. We study the observed maser decay with a composite
computational model, comprising a time-dependent chemical model of the
envelope, and a radiation transfer model which provides
the maser pumping. Our combined model is able to
reproduce the rapid decay of maser intensity, following a sudden drop in
the stellar mass-loss rate. The explanation for the rapid fall in maser 
emission is not a fall in the OH number density, or the kinetic temperature
in the inverted layers,
but the loss of a radiative
pump route which carries population from level $1$ to level $4$ via levels
$16$, $15$ and $11$. The loss of these pump routes is a result, in turn, of
a greatly reduced energy density of $53$\,$\mu$m radiation.
\end{abstract}

\section{Introduction}

When discovered in May 1988 the 1612 MHz masers of IRAS 18455+0448 had a
classic, two-horned, morphology with a peak intensity for its strongest maser
of 2.1 Jy.  However, these had faded to a solitary 110 mJy maser by July 1998,
whose subsequent exponential decline is documented by Lewis, Oppenheimer \&
Daubar \shortcite{lod01}, until it became completely undetectable in January 2001.  We thus
see that the masers of an apparently normal OH/IR star can fade by a factor of
more than a thousand in just 13 years.  Nor is 18455+0448 unique, as the 1612
MHz masers of three other OH/IR stars from the Arecibo sample have similarly
disappeared, while those of a fifth, FV Boo, are declining exponentially, and so
should disappear within the next 2-4 years \cite{murray02}.  

These five instances of `dead' OH/IR stars come from a complete sample of 328
that had $>$100 mJy masers when first detected at Arecibo.  As such they imply an
average duration, $\tau_{1612}$ , for $1612$\,MHz emission of $328 \times 14 / 5 \sim 920 (+1133,
-693)$\,yr for a general member of the population of OH/IR stars in the Arecibo
$(0\degr \leq \delta \leq 38\degr)$ sky.  But these dead stars are more particular than that as they
all have rather blue IR colours and small, $<12$\,km\,s$^{-1}$, expansion velocities.
When death statistics are calculated for the $112$ objects in the sample with
similar parameters, the maser lifetime is found to be
$\tau_{1612} = 112 \times 14 / 5 \sim 314 (+387, -97)$\,yr: the 1612 MHz
masers exhibited by this subset of OH/IR stars only have a transient existence. 

We can assimilate these observational results into our understanding of the
evolution of AGB stars, by recalling the models by Wood \& Vassiliadis \shortcite{wav92},
which  show that a brief, copious, mass-loss regime first occurs while an AGB
star is radiating away the extra energy it generates during a He-shell flash.
This fillip to the stellar luminosity lasts $\sim 500$\,yr \cite{waz81}.  But
this causality does match the brevity of the $1612$\,MHz emission phase, and does
imply that the phase will recur whenever the star passes through a He-shell
flash.  Moreover, in the case of low-mass progenitor stars, the copious mass-loss
phase is likely to dominate mass-loss while they are on the AGB, as these stars
may never be luminous enough to support much mass-loss otherwise.  The diverse
observational data supporting this scenario, which comes from studies of
high-latitude OH/IR stars, is discussed by Lewis \shortcite{murray01}.  

Until now models of the radial distribution of molecules in the circumstellar
shells of AGB stars have assumed a constant mass-loss rate over time-scales of
order $10000$\,yr (Huggins \& Glassgold \shortcite{hag82}; Netzer \& Knapp \shortcite{nak87}).  These models
are appropriate when discussing most of the classic OH/IR stars with massive
progenitors, which fall near the Galactic Plane, as these stars do eventually
reach the copious mass-loss phase while on the luminosity ascent towards a
thermal pulse.  They also have periods of order $1000-2000$\,d and expansion
velocities of $>12$\,km\,s$^{-1}$.  By contrast the low-mass and/or transient OH/IR stars
all have periods of less than $700$\,d, and usually have expansion velocities of
less than $12$\,km\,s$^{-1}$.  There is thus clearly a need to revisit the models for the
distribution of molecules about O-rich AGB stars, to incorporate the transient
aspects of their shells.  Our objective here is to determine what
properties these shells must have to reproduce the brevity of their $1612$\,MHz
masers and the speed with which normal $1612$\,MHz masers can fade away.

\section{Deriving the Radial OH Profile}

The OH found in OH/IR envelopes is assumed to be produced predominantly from
the photodissociation of H$_{2}$O. The abundance of OH at any time and position
is controlled by subsequent photodissociation of OH itself.
The UV photons in these reactions are assumed to be entirely of interstellar
origin. Two-body reverse reactions are included, though they are
endothermic and therefore relatively unimportant for the case considered in
this paper, owing to the low temperatures in the region where UV optical depth
is low enough to give significant photodissociation rates. Over
time the above  reactions lead to a classic hollow shell OH distribution 
\cite{boo81,nak87}. In order to deal with a brief superwind episode, as opposed to
continuous mass loss, a time-dependent model was constructed, involving a
series of runs, for gas emitted at different times during the superwind
episode. The ejected envelope is thus modelled as a series of concentric
spherical shells (depth elements) of thickness $v\delta t$ emitted at time
intervals $\delta t$ and drifting away from the central star at constant
terminal speed $v$ ($10$\,km\,s$^{-1}$ for the present work). The gas kinetic temperature, $T_{gas}$, is modelled by
a simple power-law function. At the time of shell detachment, the gas
temperature is given by $T_{gas} = 3.2\times 10^{14} r_{cm}^{-0.79}$, where
$r_{cm}$ is the radius in cm \cite{glassgold}.
 The change in OH number density of each depth element
over each $\delta t$ is calculated using an integration package by Gear
\shortcite{gear71}.

From the time 
when the superwind episode is terminated, no further depth
elements are produced, and those already emitted continue to drift away forming
a detached circumstellar envelope. This envelope becomes cooler and more
diffuse, causing the OH abundance to fall, as (i) dissociation rates are
rising with falling optical depth and  (ii) the H$_{2}$O from which the OH forms is
no longer being supplied to the photodissociation region.

   Though the
interstellar UV flux is assumed isotropic, the optical depth out of the
envelope (due mainly to dust) varies with direction. Therefore, in order to
estimate photodissociation rates, it is necessary to find the integral over
direction of a direction-dependent flux. In steady state models it is common
to estimate an average attenuation as a function of the outward radial
value, $\theta = 0$, for
example Morris \& Jura \shortcite{maj83}. In the case covered here we have opted for a more
accurate treatment (albeit using a very simplified density structure) due to
UV coming from previously heavily shielded directions with $\theta > \pi /2$ becoming more important as the envelope
detaches (in addition it was thought that the more accurate approach would
give extra flexibility if, for example, it were desired at some point to vary
the mass loss rate during the superwind episode instead of switching it on
then off after a certain time). The gas to dust mass ratio was set at 100 and
the UV absorption spectrum of dust taken from Massa \& Savage \shortcite{mas89}.
We note that the dust absorption spectrum used in this part of the computational
model is observationally based, and is different to the model used by the
maser pumping code (see Section~3). However, since the main spectral region
of importance in the envelope code is the ultraviolet, while the important 
region for maser pumping is the far infra-red, the two models are
not contradictory. The dust to gas mass ratio is $0.01$ in both
models.

For the
abundances adopted here, dust makes the dominant contribution to UV optical
depth, though  continuum shielding by H$_{2}$O and OH was included, using cross
section data  from van Dishoeck \& Dalgarno \shortcite{vad84}. For
details of how UV attenuation was calculated for a given column density, see
Howe \& Rawlings \shortcite{har94}. For each run, column
densities were calculated at the beginning of each time step for 6 directions
relative to the outward radial ($\theta = 0$) between $\theta = 0$ and $\theta =
\pi$, the chosen directions being dictated by the Gaussian quadrature method
used to integrate the directional flux. To avoid undue weighting of
directions with $\theta \sim \pi / 2$, the integration variable was not
$\theta$ itself, but the solid angle subtended by $\theta$.
 The total column in a given direction was
estimated by adding that of each depth element intercepted by that line of
sight, assuming each absorbing species to have a uniform fractional abundance
within that depth element, such that its number density has an inverse square
relation with distance from the central star (which can be
integrated analytically between the points where the line of sight enters and
leaves the depth element). Because the abundances of OH and H$_{2}$O are results
rather than inputs of the model, the inclusion of continuum shielding by these
species means that the model is iterated several times, starting with dust
shielding alone, until convergence is achieved. Once photorates have been
found for the beginning of each timestep, for each depth element, the rate
used to calculate abundances during the timestep was a linear interpolation
between the initial rate and that at the start of the next step (this was
easily achieved given the fact that the model was already iterative for the
shielding calculation).

\section{The OH Pumping Model}

We use a model based on the accelerated lambda iteration (ALI) method
\cite{sac85} with modifications necessary to make it applicable to
molecular line studies \cite{jones94}. The modified code, {\sc multimol},
has already been successfully applied to OH absorption \cite{jones94},
to megamaser emission \cite{rand95} and to studies of OH maser emission
in star-forming regions \cite{gray01}. Here, we combine the chemical model
described in Section~2 with {\sc multimol} to produce a time-series of
OH pumping models that follow the evolution of a circumstellar shell from
the time that the shell detaches from the host star to, at most, $1000$\,yr
later.

We adopt the slab version of {\sc multimol} to study OH, rather than the
spherical version \cite{jay98} because the former has far-infrared (FIR)
line overlap built into the code whilst the latter does not. FIR line
overlap is vital to OH pumping schemes, so we chose to model this
accurately and accept some deficiencies in the geometry, rather than the
other way about. Some quantities required geometric corrections because
the slab model was chosen; these are discussed below. The OH-containing 
shell, of which the $1612$\,MHz
maser emitting zones form a subset, is thin compared to the shell radius
in all cases except for the model with the lowest mass-loss rate. Except
in this case, the use of a slab geometry
should not introduce very significant errors into the results.

Individual slabs for each timestep were developed as follows: The chemical
model, discussed in Section~2, has linearly-spaced slabs, ordered outward
in radius from the star. However, {\sc multimol} requires logarithmically
spaced layers ordered inwards from the outer edge of the envelope. Therefore,
a service routine reversed the row order of the file, before fitting
the data with a natural cubic-spline, which was then interpolated at the
required logarithmic points, according to the formula,
\begin{equation}
z_{k} = z_{1} ( z_{M} / z_{1} )^{k/M}
\end{equation}
where $z_{k}$ is the depth of layer $k$, and there are $M$ layers altogether
in the slab. The spline and interpolation routines were taken from
Press et al. \shortcite{nrecipe}.

Other inputs to {\sc multimol} comprised a set of energy levels and
Einstein A-values for OH \cite{Destombes} and collisional rate-coefficients
for OH with molecules of ortho- and para-hydrogen \cite{off94}. A dust
model (see Section~3.1) was used to compute the continuum over the
wavelength range appropriate for the pumping lines of OH, but there was
no direct contribution from the central star.
The
models discussed here used the $36$ lowest-lying hyperfine energy levels
of OH. Competitive propagation of polarized masers involving 
magnetic hyperfine splitting \cite{gra95} was not treated here.
Following successful convergence of {\sc multimol} the solution is exact
for the geometry employed and the set of FIR transitions which form the
radiation-transfer solution. We note that, given the comments above about
the geometry, the solutions are only approximate for the spherical case.
We also note that we keep a separation of pump and maser, such that the
microwave transitions, whether masing or not, do not form part of the NLTE
radiation-transfer solution which controls the molecular
population: we only calculate unsaturated gain-coefficients
in the microwave transitions, based on inversions fixed by the FIR
transitions and collisional processes. The Monte-Carlo radiative transfer
solution by Spaans \& van Langevelde \shortcite{sav92} is more appropriate
to the geometry of the situation, and includes saturation. However, the
model in the present work includes much more detail regarding the OH molecule,
its collisions with H$_{2}$, its interaction with radiation, and the role of dust.

\subsection{The Dust Model}

The dust model used by the ALI computer code is based on theoretical
calculations of absorption and scattering efficiencies for spherical grains
by Laor and Draine \shortcite{lad93} and Draine and Lee \shortcite{dal94}. 
These efficiencies are derived from 
the tables of optical
constants in Draine \shortcite{draine85}. The optical constants were
tabulated for silicate and graphite, but for the purposes of this work, we
have made the silicate fraction overwhelming, as we are dealing with
oxygen-rich stars, which we assume do not produce carbon dust. At all
modelled points in the envelope, we assume a constant dust mass fraction of
1\%. The size of dust grains was allowed to vary from $1$\,nm up to
$10$\,$\mu$m, following a power-law spectral index equal to $-3.5$. In this
respect, the model resembles MRN \cite{mrn77} dust, with an extended size range
and no carbon component. For the spectra of several MRN-like models, and
comparison with other theoretical and observational dust `laws', see
Gray \shortcite{gray01}. In particular, the parameters of the dust model
in the present work are very similar to the model `E' in Gray
\shortcite{gray01}, but without the graphite.

The dust model described here was only used to provide a continuum in the
spectral vicinity of the OH pumping lines (from about $10$ to 
$200$\,$\mu$m) with most of these
lines lying in the far infra-red. The dust absorption spectrum was not
used to compute dynamical effects, via radiation pressure, nor to calculate
dust temperatures. The dust temperature was instead related to the gas
kinetic temperature via simple empirical formulae, which could be changed
from model to model (see Section~4). In most cases the formula used was
$T_{dust}=max(25,T_{gas}-25)$, where $T_{dust}$ and $T_{gas}$ are the dust and
gas temperatures in Kelvin. The minimum dust temperature was used to represent
the value which can be maintained by typical interstellar irradiation
\cite{spitz78}. 

\section{Results}

\begin{figure}
\vspace{0.3cm}
\psfig{file=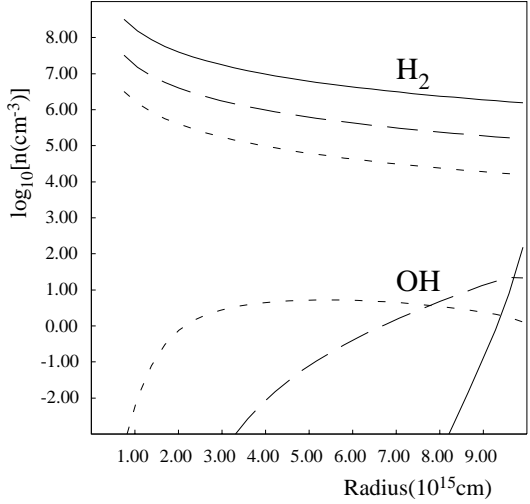,angle=0,width=8cm}
\caption{
The variation, at shell detachment, of the number densities of H$_{\rm 2}$ and
OH, as a function of radius in the shell. Data are plotted for three mass-loss
rates: $1.0\times 10^{-4}$\,M$_{\odot}$\,yr$^{-1}$ (solid line),
$1.0\times 10^{-5}$\,M$_{\odot}$\,yr$^{-1}$ (long dashes) and
$1.0\times 10^{-6}$\,M$_{\odot}$\,yr$^{-1}$ (short dashes).
}
\label{f:figzero}
\end{figure}

Our model, comprising the hydrodynamic and photo-chemical model of the
detached envelope (see Section~2) and the radiation transfer code discussed
in Section~3, was run for a series of stellar mass-loss rates. These rates
refer to values of $\dot{M}$ prior to a time when the envelope detaches,
when the mass-loss rate is assumed to fall to a negligibly low value. Seven
versions of the model were run, with mass-loss rates, $\dot{M}$, ranging from
$1.0\times 10^{-6}$\,M$_{\odot}$\,yr$^{-1}$ up to $1.0\times 10^{-4}$\,M$_{\odot}$\,yr$^{-1}$. We show the distribution of the
number densities of molecular
hydrogen and OH as a function of radius in the shell in Figure~\ref{f:figzero}. Graphs are
plotted for three mass-loss rates, including the lowest and highest used. All
data are plotted at the time of shell detachment.
An additional model was run at a mass-loss
rate of $1.0\times 10^{-4}$\,M$_{\odot}$\,yr$^{-1}$, but a longer ($600$\,yr)
elapsed time before shell detachment. The elapsed time was $300$\,yr in all
other cases.

We note that the model run at a mass-loss rate of
$1.0\times 10^{-6}$\,M$_{\odot}$\,yr$^{-1}$ has a significant abundance of
OH deep inside the envelope. The accuracy of the slab model as an
approximation to spherical geometry is therefore poor in this case, and to
a lesser extent in the model at 
$3.0\times 10^{-6}$\,M$_{\odot}$\,yr$^{-1}$. Accuracy of the geometrical
representation rises rapidly with mass-loss rate. 

In each model the time begins at zero and, prior to shell detachment, the
model star loses mass at the rate applicable to that version. At shell
detachment, usually $t=300$\,yr,
the mass-loss rate falls to zero, and the circumstellar shell becomes
progressively more hollow.
This corresponds to the end of a superwind phase in the language
of Section~2.
From detachment onwards, the unsaturated maser gain is calculated
at each timestep. The calculations were continued until the gain at $1612$\,MHz
had clearly fallen into absorption (indicated here, 
and in Tables~1-7, by {\it negative} values of
$\Gamma$). This only took more than $100$\,yr
in the one-off model in which shell detachment occurred at $600$\,yr.
Detached solutions were available to continue calculations
until $t=1500$\,yr, had this been necessary. In Tables~1-7, we show the
variation of the integrated gain through the shell in the four ground-state
lines of OH as a function of time for seven mass-loss rates and
shell detachment at $300$\,yr.
Times prior to shell detachment are not shown. Where information for times is missing, the
radiative transfer code did not reach an acceptable solution within
a reasonable number ($250$) of iterations.

\begin{table}
\caption{$\dot{M}=1.0\times 10^{-6}$\,M$_{\odot}$\,yr$^{-1}$}
\begin{tabular}{@{}lrrrr@{}}
\hline
Time   &  $\Gamma_{1720}$& $\Gamma_{1667}$& $\Gamma_{1665}$& $\Gamma_{1612}$\\
 (yr)  &                 &                &                &                \\
\hline
300    &    -19.827      &   -84.607      &   -36.591      &   3.107        \\
310    &    -16.127      &   -92.746      &   -41.219      &  -2.421        \\
320    &    -15.085      &  -102.321      &   -46.477      &  -5.579        \\
340    &    -16.421      &  -122.991      &   -61.982      &  -9.640        \\ 
360    &    -15.170      &  -131.971      &   -66.790      & -12.894        \\
380    &    -12.656      &  -133.398      &   -68.372      & -15.838        \\
\hline
\end{tabular}
\end{table}

\begin{table}
\caption{$\dot{M}=3.0\times 10^{-6}$\,M$_{\odot}$\,yr$^{-1}$}
\begin{tabular}{@{}lrrrr@{}}
\hline
Time   &  $\Gamma_{1720}$& $\Gamma_{1667}$& $\Gamma_{1665}$& $\Gamma_{1612}$\\
 (yr)  &                 &                &                &                \\
\hline
300    &    -26.345      &   -62.275      &   -22.615      &  14.900        \\
310    &    -27.539      &   -87.366      &   -42.524      &   9.325        \\
320    &    -28.104      &  -100.424      &   -53.198      &   6.306        \\
330    &    -26.448      &  -109.520      &   -59.880      &   2.303        \\
340    &    -25.634      &  -117.503      &   -65.666      &  -0.555        \\ 
360    &    -24.903      &  -132.800      &   -75.380      &  -4.928        \\
370    &    -25.061      &  -140.388      &   -79.799      &  -6.496        \\
380    &    -24.657      &  -146.187      &   -84.244      &  -8.433        \\
390    &    -23.798      &  -152.630      &   -88.136      & -10.786        \\
\hline
\end{tabular}
\end{table}

\begin{table}
\caption{$\dot{M}=1.0\times 10^{-5}$\,M$_{\odot}$\,yr$^{-1}$}
\begin{tabular}{@{}lrrrr@{}}
\hline
Time   &  $\Gamma_{1720}$& $\Gamma_{1667}$& $\Gamma_{1665}$& $\Gamma_{1612}$\\
 (yr)  &                 &                &                &                \\
\hline
300    &    -11.324      &     2.603      &    16.831      &  14.977        \\
310    &    -17.383      &   -24.675      &    -5.892      &  13.461        \\
320    &    -20.417      &   -43.418      &   -21.057      &  11.380        \\
330    &    -22.290      &   -58.811      &   -32.998      &   9.155        \\
340    &    -23.219      &   -70.508      &   -41.848      &   7.014        \\
350    &    -23.640      &   -81.343      &   -49.657      &   4.670        \\
360    &    -23.847      &   -91.470      &   -56.723      &   2.339        \\
370    &    -24.051      &  -100.989      &   -63.243      &  -0.002        \\
\hline
\end{tabular}
\end{table}

\begin{table}
\caption{$\dot{M}=1.5\times 10^{-5}$\,M$_{\odot}$\,yr$^{-1}$}
\begin{tabular}{@{}lrrrr@{}}
\hline
Time   &  $\Gamma_{1720}$& $\Gamma_{1667}$& $\Gamma_{1665}$& $\Gamma_{1612}$\\
 (yr)  &                 &                &                &                \\
\hline
300    &      1.690      &    36.579      &    35.366      &   9.446        \\
310    &     -3.956      &    16.069      &    19.131      &   9.567        \\
320    &     -7.784      &    -0.829      &     5.779      &   8.846        \\
330    &    -10.685      &   -16.034      &    -5.735      &   7.755        \\
340    &    -12.537      &   -28.310      &   -14.824      &   6.426        \\
350    &    -14.086      &   -40.842      &   -23.576      &   4.832        \\
360    &    -15.044      &   -52.797      &   -31.550      &   2.867        \\
370    &    -15.613      &   -62.601      &   -37.830      &   1.091        \\
380    &    -15.869      &   -74.144      &   -44.734      &  -1.316        \\
390    &    -15.581      &   -85.750      &   -51.234      &  -4.193        \\
400    &    -15.361      &   -94.955      &   -56.277      &  -6.444        \\
\hline
\end{tabular}
\end{table}

\begin{table}
\caption{$\dot{M}=2.0\times 10^{-5}$\,M$_{\odot}$\,yr$^{-1}$}
\begin{tabular}{@{}lrrrr@{}}
\hline
Time   &  $\Gamma_{1720}$& $\Gamma_{1667}$& $\Gamma_{1665}$& $\Gamma_{1612}$\\
 (yr)  &                 &                &                &                \\
\hline
300    &      3.998      &    34.676      &    32.683      &   6.391        \\
310    &     -0.718      &    19.678      &    20.868      &   7.077        \\
320    &     -4.279      &     5.940      &     9.924      &   6.902        \\
330    &     -7.243      &    -7.214      &    -0.307      &   6.379        \\
340    &     -9.240      &   -18.131      &    -8.663      &   5.492        \\
350    &    -11.114      &   -29.774      &   -17.060      &   4.393        \\
360    &    -12.407      &   -41.079      &   -24.903      &   2.862        \\
370    &    -13.285      &   -50.386      &   -31.138      &   1.459        \\
380    &    -13.968      &   -61.842      &   -38.314      &  -0.566        \\
390    &    -14.059      &   -73.284      &   -44.983      &  -3.080        \\
400    &    -14.168      &   -82.713      &   -50.375      &  -5.096        \\
\hline
\end{tabular}
\end{table}

\begin{table}
\caption{$\dot{M}=3.0\times 10^{-5}$\,M$_{\odot}$\,yr$^{-1}$}
\begin{tabular}{@{}lrrrr@{}}
\hline
Time   &  $\Gamma_{1720}$& $\Gamma_{1667}$& $\Gamma_{1665}$& $\Gamma_{1612}$\\
 (yr)  &                 &                &                &                \\
\hline
300    &      6.027      &    21.353      &    23.724      &   1.090        \\
310    &      2.081      &    12.316      &    16.850      &   2.657        \\  
320    &     -1.427      &     2.831      &     8.952      &   3.531        \\
330    &     -4.511      &    -7.737      &     0.583      &   3.749        \\
340    &     -6.849      &   -17.022      &    -7.097      &   3.537        \\
350    &     -8.912      &   -26.159      &   -14.169      &   3.172        \\
360    &    -10.791      &   -37.057      &   -22.457      &   2.182        \\
370    &    -12.112      &   -45.578      &   -28.792      &   1.289        \\
380    &    -13.307      &   -54.540      &   -35.215      &   0.204        \\
390    &    -14.207      &   -66.223      &   -43.199      &  -1.790        \\
400    &    -14.869      &   -75.468      &   -49.340      &  -3.383        \\
\hline
\end{tabular}
\end{table}

\begin{table}
\caption{$\dot{M}=1.0\times 10^{-4}$\,M$_{\odot}$\,yr$^{-1}$}
\begin{tabular}{@{}lrrrr@{}}
\hline
Time   &  $\Gamma_{1720}$& $\Gamma_{1667}$& $\Gamma_{1665}$& $\Gamma_{1612}$\\
 (yr)  &                 &                &                &                \\
\hline
300    &      3.893      &     2.485      &     4.595      &  -2.698        \\
310    &      4.424      &     1.628      &     4.787      &  -3.285        \\
320    &      4.115      &     1.182      &     4.530      &  -3.077        \\
330    &      3.827      &    -1.299      &     3.324      &  -3.307        \\
340    &      2.593      &    -3.071      &     1.424      &  -2.649        \\
350    &      1.966      &    -6.396      &    -0.580      &  -2.792        \\
360    &      0.328      &   -10.805      &    -4.382      &  -2.405        \\
370    &     -0.777      &   -14.527      &    -7.335      &  -2.303        \\
\hline
\end{tabular}
\end{table}

The general trend from the tables is that the gain of the $1612$\,MHz masers decays to
absorption on a timescale of a few decades after shell detachment. Apart from
the case of the highest mass-loss rate (Table~7) there are always $1612$\,MHz
masers present at the time of shell detachment. The extra model, with shell
detachment at $600$\,yr, does have $1612$\,MHz masers present at the time
of detachment: in this case the initial integrated gain of $1.728$, rises
to a maximum of $5.211$ at $40$\,yr after detachment, followed by a decay
to absorption lasting an additional $80$\,yr. For the models with
shell detachment at $300$\,yr,
the initial masers are
strongest between mass-loss rates of $3.0\times 10^{-6}$\,M$_{\odot}$\,yr$^{-1}$ and
$1.0\times 10^{-5}$\,M$_{\odot}$\,yr$^{-1}$. At both these values of $\dot{M}$ the
maser intensity at $t=300$\,yr would in practice be limited by saturation.
The longest-lasting $1612$\,MHz masers are at $\dot{M}=3.0\times 10^{-5}$\,M$_{\odot}$\,yr$^{-1}$, where
they remain inverted until $80$\,yr after
shell detachment. Except at $\dot{M}=1.0\times 10^{-4}$\,M$_{\odot}$\,yr$^{-1}$, any
masers initially present in the other three lines decay substantially more
quickly than those at $1612$\,MHz. However, for the main lines, the current
model ignores masers generated deep within the shell (in a zone shared with
water masers, for example Richards et al. \shortcite{anita02}) where there would be a shock source of OH, rather than the
source generated by interstellar UV photodissociation which is used in the
present work.

We proceed to compare the decay times of our model $1612$\,MHz masers with
the values obtained by Lewis et al. \shortcite{lod01} for IRAS 18455+0448,
and by Lewis \shortcite{murray02} for FV~Boo.

\subsection{Model Parameters and Decay Times}

Here we discuss the input parameters for the models discussed in
Section~4. Parameters that were considered
standard for all models
are set out in Table~8. The thickness of the OH shell is the maximum
found, corresponding to the smallest mass-loss rate of
10$^{-6}$\,M$_{\odot}$\,yr$^{-1}$. Typical values at higher mass-loss
rates were considerably smaller.  
In all the models that were run to produce the data in Tables~$1$-$7$, the dust temperature 
in the outer envelope was not allowed to fall below $25$\,K, a
value maintained by the interstellar radiation field \cite{spitz78}. An additional test model was also run in which the dust temperature
was linked only to the gas temperature, being offset from it by $-25$\,K, with
no minimum other than absolute zero.
At a mass-loss rate of 10$^{-5}$\,M$_{\odot}$\,yr$^{-1}$, differences
between the test model and the data in Table~3 were less than 5\% in the
worst case. 

\begin{figure*}
\vspace{0.3cm}
\psfig{file=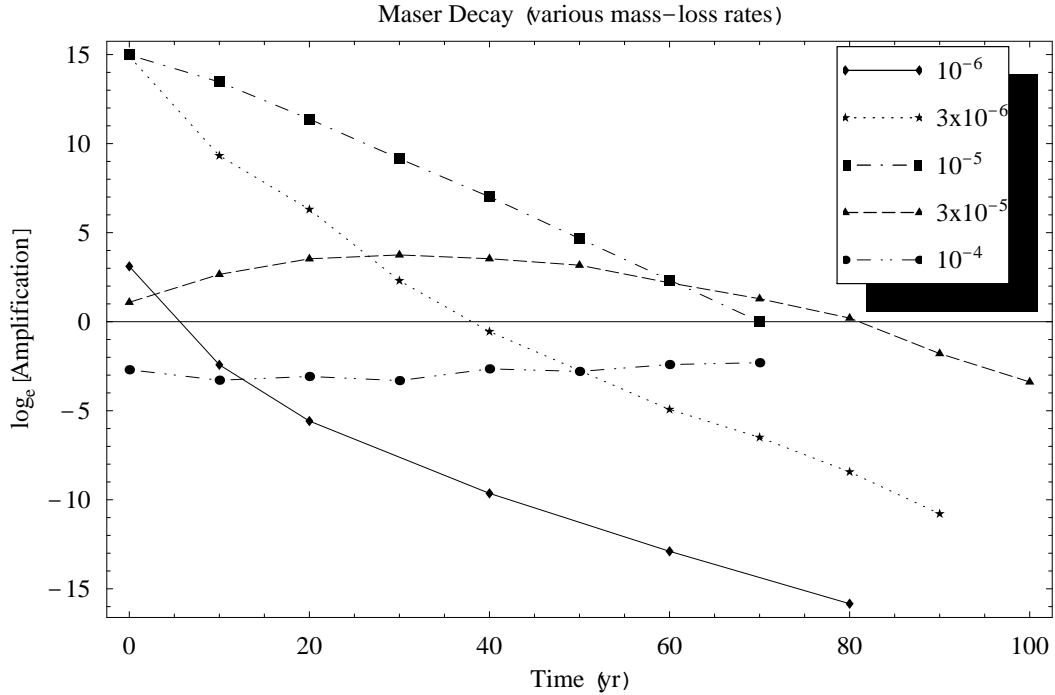,angle=0,width=16cm}
\caption{
Decay of the line-centre maser gain as a function of time for model stars
with five different mass-loss rates (in solar masses per year). Note that
the amplifications given do not take saturation into account. Values of the
logarithm above $\sim$10-11 would be about the maximum achieved under
the limit of strong saturation,
}
\label{f:figone}
\end{figure*}

For the models in Tables~$1$-$3$, $6$ and $7$, the
value of the integrated gain coefficient at line-centre for the $1612$\,MHz
maser, as a function of time, is plotted in Fig.~\ref{f:figone}. The time origin is
taken as the moment of shell detachment.

\begin{table}
\caption{Model Parameters}
\begin{tabular}{@{}lrr@{}}
\hline
Parameter              &  Value &  Unit \\
\hline
Shell depth            & $8.13 \times 10^{15}$   &  cm     \\
Thickness of OH shell (max.) & $7.21 \times 10^{15}$   &  cm     \\
Number of depth points &           82            &         \\
OH levels modelled     &           36            &         \\
Radiation angles       &            3            &         \\
Velocity shift         &          0.0            & km\,s$^{-1}$ \\
Microturbulent velocity&          0.0            & km\,s$^{-1}$ \\
Expansion velocity     &           10            & km\,s$^{-1}$ \\
Dust temperatures      & $max(T_{gas}-25,25)$    & K \\
\hline
\end{tabular}
\end{table}

For each of the models which has a strong $1612$-MHz maser at the time of
shell detachment, we compute a decay time, over which the unsaturated
maser amplification falls to $1/2$ of its original value. The models
considered in this way comprise Tables~$1$-$6$. We note that in
model $6$ there is a period of initial increase in amplification factor. In 
this case, we add the time to reach maximum ($30$\,yr) to the decay time.

\begin{figure*}
\vspace{0.3cm}
\psfig{file=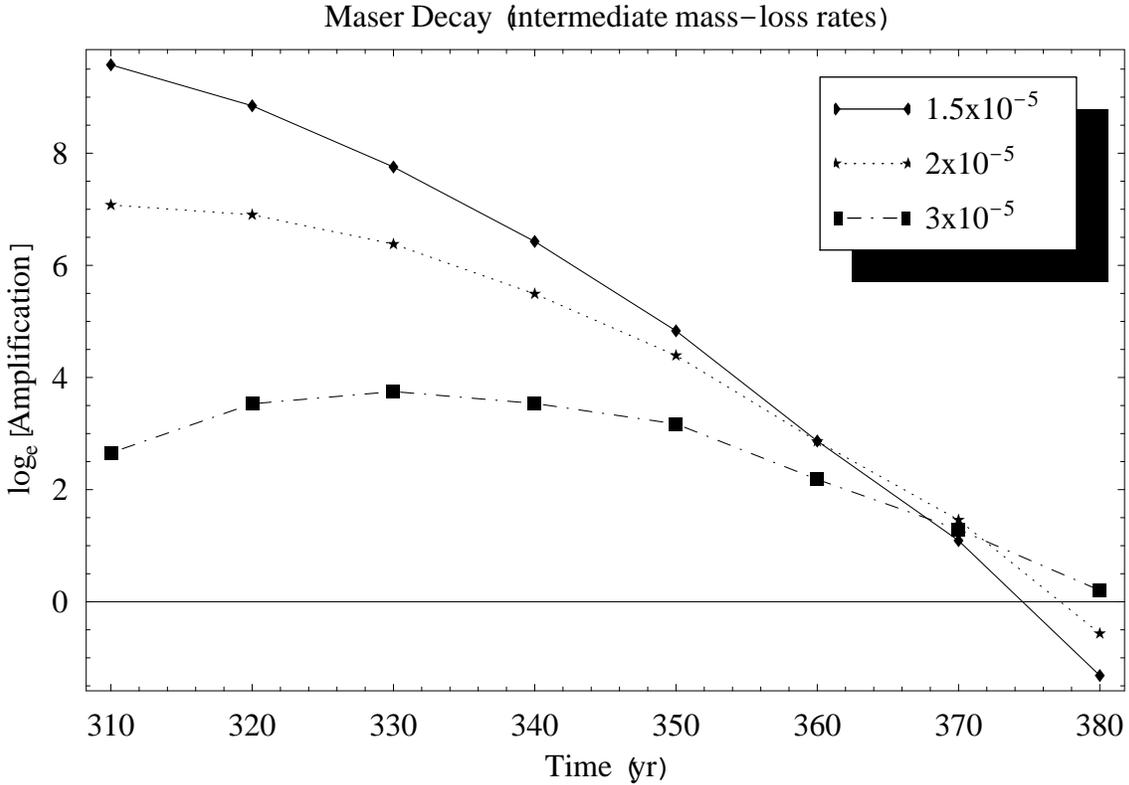,angle=0}
\caption{
Decay curves for three models at mass loss rates between 
$10^{-5}$\,M$_{\odot}$\,yr$^{-1}$ and $10^{-4}$\,M$_{\odot}$\,yr$^{-1}$,
showing the initial slow decay from the peak of each curve, followed by
an approximately linear fall with time.
}
\label{f:figtwo}
\end{figure*}

In all cases, we assume that changes in the observed flux density of the
$1612$-MHz maser result from a change in the pumping scheme, which reduces
the maser gain coefficient. The observed fluxes are then proportional to
total amplification factors in the maser column. We ignore saturation here,
though the masers in Tables~$2$ and $3$ would probably be partially saturated.
Suppose we want to calculate the time taken for an observed flux density to
fall to some fraction, $r$, of its value at the time of shell detachment.
The fraction is given by
\begin{equation}
r = \frac{S_{\nu}(t)}{S_{\nu}(t_{0})} = \frac{A(t)}{A(t_{0})}
\end{equation}
where $S_{v}(t)$ is a flux density, and $A(t)$ is a computed amplification
factor. The amplification is given in terms of the overall integrated
gain, $\Gamma$,
of the model as $A(t) = \exp ( \Gamma (t) )$. Therefore the logarithm of
$r$ is just given by the difference in the integrated gains at the two
times, or
\begin{equation}
\ln r = \Gamma (t) - \Gamma (t_{0})
\end{equation}
For small times, $t-t_{0}$, we can expand $\Gamma (t)$ in a Taylor series
about $t=t_{0}$, and truncate at the linear term, leaving
\begin{equation}
\Gamma (t) \sim \Gamma (t_{0}) + \left. \frac{d \Gamma}{dt} \right|_{t=t_{0}}
(t - t_{0}) 
\end{equation}
This can be used in eq.(3) to eliminate $\Gamma(t_{0})$, and supposing that
we choose $r=1/2$, we can obtain an estimate of the `half-life' of the
maser decay as,
\begin{equation}
\tau_{1/2} = (t-t_{0})_{1/2} = \frac{-\ln 2}{(d\Gamma / dt)|_{t=t_{0}}}
\end{equation}
If we approximate the differential as the finite difference between the
initial time and the second computation ($10$\,yr later), we find
half-lives for maser decay of $1.25$\,yr at a mass-loss rate of 
$10^{-6}$\,M$_{\odot}$\,yr$^{-1}$, falling slightly to $1.24$\,yr at
$3.0\times 10^{-6}$\,M$_{\odot}$\,yr$^{-1}$, and then rising to
$4.57$\,yr when the mass-loss rate reaches $10^{-5}$\,M$_{\odot}$\,yr$^{-1}$.
For mass-loss rates larger than $10^{-5}$\,M$_{\odot}$\,yr$^{-1}$, the
situation is complicated by an initial period of growth in maser amplification
before the decay sets in. A naive method of calculating the decay time in
these cases is to take the decay time from the peak. For example, at
$3\times 10^{-5}$\,M$_{\odot}$\,yr$^{-1}$, the peak is reached after
$30$\,yr and, starting from this time, the half-life from eq.(5) is 
some $33$\,yr. A more sophisticated analysis is to view the data in
Tables~$4$-$6$ as an intermediate case
between the $10^{-4}$\,M$_{\odot}$\,yr$^{-1}$ shell, where not
enough time has elapsed before shell detachment
for $1612$\,MHz OH masers to form, and the lower
mass-loss rate cases, where decay of these masers begins immediately after
shell detachment. The extra model in which the shell detaches after $600$\,yr,
as discussed in Section~4, shows that $1612$\,MHz masers can form in
envelopes with loss-rates of $10^{-4}$\,M$_{\odot}$\,yr$^{-1}$, given a
longer episode of mass loss before detachment.
It is therefore instructive to plot the decay of the
$1612$\,MHz masers in the three cases which have mass loss rates 
between $10^{-5}$\,M$_{\odot}$\,yr$^{-1}$ and
$10^{-4}$\,M$_{\odot}$\,yr$^{-1}$ without the other models present (Fig.~\ref{f:figtwo}).
 We can see from Fig.~\ref{f:figtwo} that
an inital slow decline from the peak, in all three cases,
is followed
by a roughly linear fall in $\Gamma$ with time. Basing the decay of
the $1612$-MHz masers in the $3$ intermediate models on
this linear section of the graph only, we find a much shorter half-life
of $7.01$\,yr for the $3\times 10^{-5}$\,M$_{\odot}$\,yr$^{-1}$ case. The
linear section is taken to start twenty years after the peak, and to continue
until the last positive value of $\Gamma$.
Similar calculations yield $4.16$\,yr for $1.5\times 10^{-5}$\,M$_{\odot}$\,yr$^{-1}$
and $5.64$\,yr for $2\times 10^{-5}$\,M$_{\odot}$\,yr$^{-1}$. We note that
the models with mass-loss rates of $10^{-5}$\,M$_{\odot}$\,yr$^{-1}$ and
lower also exhibit
quasi-linear declines of $\Gamma$ with time.
These computational results compare to a decay time for the red peak
of FV~Boo of some $250-300$\,d ($0.68-0.82$\,yr).

\begin{figure*}
\vspace{0.3cm}
\psfig{file=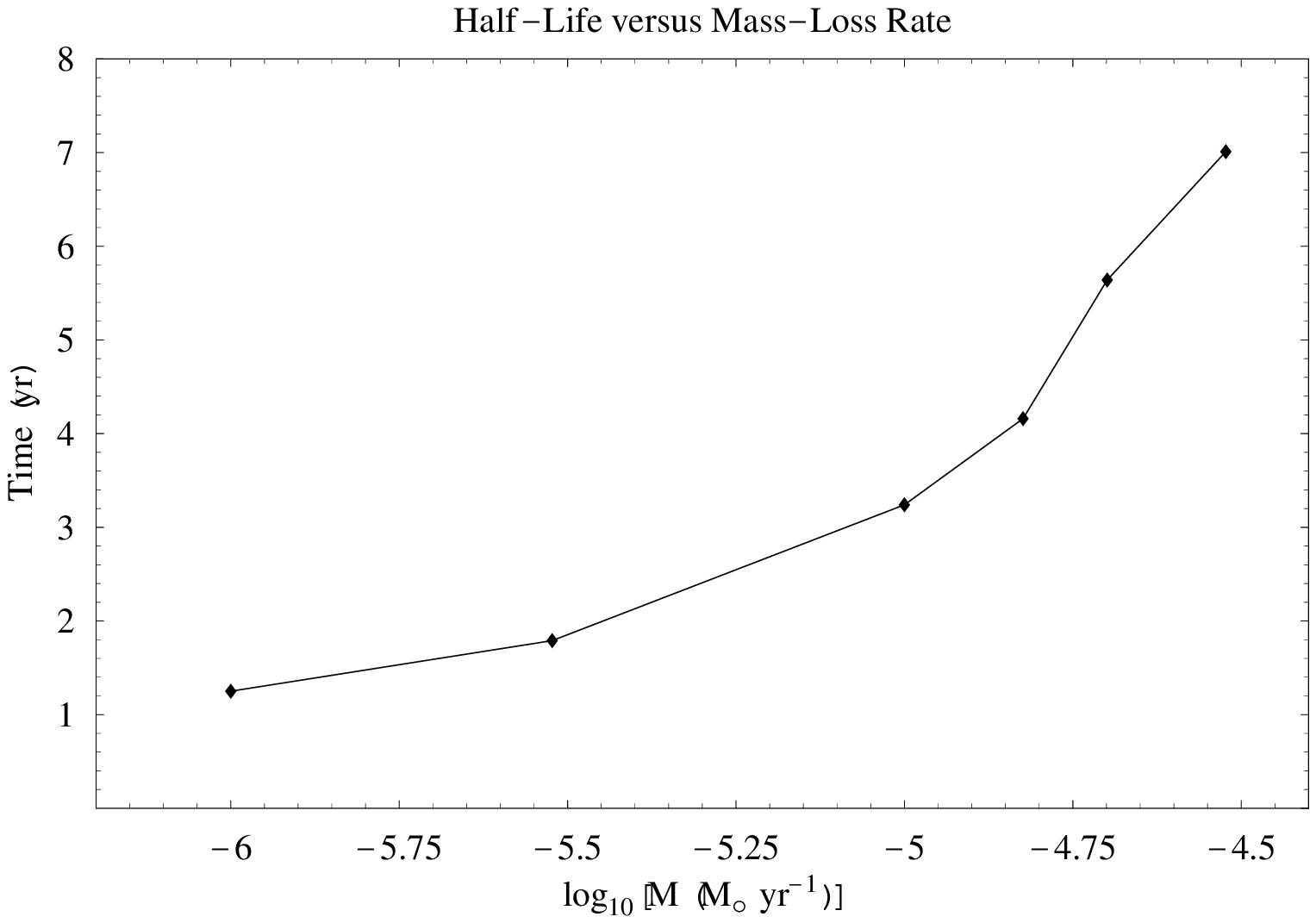,angle=0}
\caption{
The variation of the decay half-life of $1612$\,MHz masers as a function of
mass-loss rate.
}
\label{f:figthree}
\end{figure*}

We go on to consider the variation of the maser decay half-life,
$\tau_{1/2}$, as a function of the mass-loss rate. For the mass-loss rates
above $10^{-5}$\,M$_{\odot}$\,yr$^{-1}$, we use the values of the half-life
computed from the quasi-linear region of Fig.~\ref{f:figtwo}. The relationship is
plotted in Fig.~\ref{f:figthree}.
For the remaining points, we compute half-lives based
on all points in Fig.~\ref{f:figone} up to, and including, the first point with a negative
value of $\Gamma$. This rule yields values of $\tau_{1/2}$, for the mass
loss rates of $1$, $3$ and $10\times 10^{-6}$\,M$_{\odot}$\,yr$^{-1}$, of
$1.25$, $1.79$ and $3.24$\,yr, respectively.
For the models with mass-loss rates of $10^{-5}$\,M$_{\odot}$\,yr$^{-1}$
and lower, we find a linear relation between the half-life of decay
and the mass-loss rate, which is
\begin{equation}
\tau_{1/2} = 13.19 + 1.99 \log_{10}(\dot{M}/(M_{\odot}\,yr^{-1})) \; \; {\rm yr}
\end{equation}
A possible explanation for the change in the behaviour of the half-life at a
mass-loss rate of $\sim 10^{-5}$\,M$_{\odot}$\,yr$^{-1}$ is that $300$\,yr is
sufficient for masers to become saturating for mass-loss rates lower than
this, and at higher mass-loss rates the shell has not been undergoing
a superwind loss for long enough to achieve saturation. This view is
supported by the appearance of strong masers in the model run after
$600$\,yr of mass-loss for a loss rate of $10^{-4}$\,M$_{\odot}$\,yr$^{-1}$.

\subsection{Position of Emitting Shells}

Here we consider the zones in the envelope which contribute most to the maser
output for Tables $1$-$7$. At the time of shell detachment, and for
most of the models, the OH that
sustains an inversion is divided into two layers: a very thin layer bordering
the outer edge of the shell, and a deeper layer which is much thicker,
although the inversions are smaller. 
As the mass-loss rate increases, the
already thin
outer layer becomes physically even thinner, though its inverted column density
increases somewhat,
whilst the deeper layer gets closer to the
surface, and narrows. A very weakly inverted third emission layer develops
for mass-loss rates above $3.0 \times 10^{-5}$\,M$_{\odot}$\,yr$^{-1}$, but
this contributes little to the maser output of the star. For all models, it
is the second inverted layer which contributes the great majority of the maser output.
The absorbing layer, found between this and the outer emission zone, can
absorb anything from $\sim 10$ per cent to $100$ per cent of the maser
emission. The outer emitting layer typically adds only a few per cent or
less of the maser emission generated by the deeper emitting zone.

\begin{table*}
\begin{minipage}{16cm}
\caption{Parameters of emitting and absorbing zones of the envelope at the time
of shell detachment. Column densities given are the column densities per
sublevel of
OH molecules in the upper level of the 1612\,MHz transition (level $3$) minus
the corresponding column densities of OH, per sublevel, in the lower level of this transition (level $2$). This definition means that
inverted column densities
are positive; absorbing column densities are negative. The terms Zone~1 and Zone~2
refer to the outer and inner inverted layers respectively. Depths, $z$, are
measured from the outer (ISM) surface of the envelope. $z_{out}$ for Zone~1 is
zero in all cases.}
\begin{tabular}{@{}lrrrrrr@{}}
\hline
$\dot{M}$             &  $z_{in}$ Zone~1   & $z_{out}$ Zone~2   & $z_{in}$ Zone~2    & Zone~1 column      & Interzone absorption&  Zone~2 column    \\
M$_{\odot}$\,yr$^{-1}$&      cm            &      cm            &       cm           &     cm$^{-2}$      &   cm$^{-2}$         &      cm$^{-2}$    \\
\hline
$1.0\times 10^{-6}$  &$9.98\times 10^{12}$&$1.52\times 10^{15}$&$6.40\times 10^{15}$&$7.35\times 10^{10}$&$-8.39\times 10^{12}$&$2.76\times 10^{13}$\\
$3.0\times 10^{-6}$  &$4.32\times 10^{12}$&$2.24\times 10^{14}$&$4.47\times 10^{15}$&$1.75\times 10^{11}$&$-4.27\times 10^{12}$&$7.15\times 10^{13}$\\
$1.0\times 10^{-5}$  &$1.47\times 10^{12}$&$3.72\times 10^{13}$&$2.18\times 10^{15}$&$3.13\times 10^{11}$&$-2.94\times 10^{12}$&$6.23\times 10^{13}$\\
$1.5\times 10^{-5}$  &$1.03\times 10^{12}$&$1.82\times 10^{13}$&$1.06\times
10^{15}$&$3.84\times 10^{11}$&$-1.47\times 10^{12}$&$3.42\times 10^{13}$\\
$2.0\times 10^{-5}$  &$9.10\times 10^{11}$&$1.27\times 10^{13}$&$6.58\times
10^{14}$&$4.58\times 10^{11}$&$-2.42\times 10^{12}$&$2.27\times 10^{13}$\\
$3.0\times 10^{-5}$  &$8.08\times 10^{11}$&$1.27\times 10^{13}$&$3.21\times 10^{14}$&$5.59\times 10^{11}$&$-2.88\times 10^{12}$&$9.26\times 10^{12}$\\
$1.0\times 10^{-4}$  &$7.17\times 10^{11}$&$7.85\times 10^{12}$&$2.93\times 10^{13}$&$6.78\times 10^{11}$&$-2.80\times 10^{12}$&$8.25\times 10^{11}$\\
\hline
\end{tabular}
\end{minipage}
\end{table*}

\begin{table*}
\begin{minipage}{16cm}
\caption{As for Table~9, but for a time of $40$\,yr after shell detachment. Note that for the lowest mass-loss rate, the deeper inverted
layer has entirely disappeared, and absorption persists to the maximum depth of the model.}
\begin{tabular}{@{}lrrrrrr@{}}
\hline
$\dot{M}$             &  $z_{in}$ Zone~1   & $z_{out}$ Zone~2   & $z_{in}$ Zone~2    & Zone~1 column      & Interzone absorption&  Zone~2 column    \\
M$_{\odot}$\,yr$^{-1}$&      cm            &      cm            &       cm           &     cm$^{-2}$      &   cm$^{-2}$         &      cm$^{-2}$    \\
\hline
$1.0\times 10^{-6}$  &$1.82\times 10^{13}$&  N/A               &   N/A              &$8.19\times 10^{10}$&       N/A            &      N/A          \\
$3.0\times 10^{-6}$  &$6.18\times 10^{12}$&$2.77\times 10^{15}$&$7.21\times 10^{15}$&$1.55\times 10^{11}$&$-1.37\times 10^{13}$&$1.86\times 10^{13}$\\
$1.0\times 10^{-5}$  &$1.87\times 10^{12}$&$1.99\times 10^{14}$&$5.04\times 10^{15}$&$2.78\times 10^{11}$&$-6.80\times 10^{12}$&$3.67\times 10^{13}$\\
$1.5\times 10^{-5}$  &$1.16\times 10^{12}$&$5.33\times 10^{13}$&$2.77\times
10^{15}$&$3.06\times 10^{11}$&$-4.41\times 10^{12}$&$2.94\times 10^{13}$\\
$2.0\times 10^{-5}$  &$1.03\times 10^{12}$&$3.30\times 10^{13}$&$2.18\times
10^{15}$&$3.73\times 10^{11}$&$-4.10\times 10^{12}$&$2.51\times 10^{13}$\\
$3.0\times 10^{-5}$  &$9.10\times 10^{11}$&$3.30\times 10^{13}$&$2.18\times 10^{15}$&$5.27\times 10^{11}$&$-4.25\times 10^{12}$&$1.83\times 10^{13}$\\
$1.0\times 10^{-4}$  &$7.17\times 10^{11}$&$1.12\times 10^{13}$&$4.19\times 10^{13}$&$6.86\times 10^{11}$&$-3.17\times 10^{12}$&$6.47\times 10^{11}$\\
\hline
\end{tabular}
\end{minipage}
\end{table*}

In Table~9, we summarise the parameters of the two zones with OH inversions,
and the absorbing layer which falls between them. Table~9 shows the state
of the envelope at the time of shell detachment. To show how the inversion
zones develop after the cessation of rapid mass-loss, we present, in Table~10,
the emission zone parameters forty years later. The dominant effect which, at
this level, explains the decreasing maser intensity discussed in Section~4.1,
is that the inner inverted zone moves away from the surface of the envelope,
and suffers a steadily decreasing inverted column. The next most important
effect is that the absorbing column external to the deep inverted layer
increases with time following shell detachment.

\section{Population Tracing}

Here, we present a method of recovering the dominant routes for population
transfer that maintain inversions in the $1612$\,MHz line. As the method of
analysis becomes very lengthy in cases where the population transfer network
is complicated, we consider just the case of a mass-loss rate
of $10^{-5}$\,M$_{\odot}$\,yr$^{-1}$ (see Table~3). Within this one model,
we study how the population transfer routes vary with position in the
envelope at a fixed time: we analyse the differences between the main
inverted layer (zone~2 in Tables~9 and 10), the thin outer inverted layer
and the intermediate absorbing layer. We also consider the time evolution of
zone~2, comparing the population transfer routes at shell detachment with
those $40$\,yr later. The variations with position are considered only
at shell detachment.

\subsection{The Computer Code {\sc tracer}}

The ALI code {\sc multimol}, used to compute all the preceeding numerical results for this
work, has at its core a linear algebra solver. In this respect, it is typical
of most radiation transfer codes. The usual requirement is to reduce a large
matrix of coefficients
to upper echelon form, and this is commonly done via a stable 
numerical method, such as Gauss elimination or LU-factorisation of the matrix.
{\sc multimol} uses LU-factorisation. The problem with using such techniques is that,
although a numerical result is achieved, the information about how population
is actually transferred through the energy levels of the molecule is lost.
The computer code {\sc tracer} restores information about population transfer
routes by taking a converged ALI solution, and re-computing it via the
naive `schoolchild algebra' technique. This method can be used to trace
the development of each rate-coefficient as the matrix is modified, and
has the advantage of a simple physical interpretation.

Suppose that we take one slab of the ALI solution. Populations of energy
levels in the slab are decided by a set of equations that describe the
population flow into and out of each level. One equation, the one for the
ground state, is replaced by a conservation equation in order to make the
system of equations inhomogeneous, and to aid numerical stability. However,
it is easy to maintain a parallel set of coefficients for the `real'
ground-state equation for tracing processes. This set will be used to trace
all coefficients for population going into level $1$, but it is ignored
in favour of the conservation equation for
the purposes of actually calculating level populations.  

In all the other
equations, the diagonal coefficient represents the flow out of a given
energy level, and all the other coefficients represent the flow into it from
all the other levels; in a steady-state, the inward and outward contributions
sum to zero. Take, as an example, a situation where sixteen eliminations
remain to be carried out in the matrix. We will write the all-process rate
coefficient for transfer of population from level $2$ to level $10$, at this
stage of the elimination process, as $k_{2,10}^{16}$. When the next 
elimination is carried out, this rate-coefficient is modified to
become $k_{2,10}^{15}$, and from the `schoolchild algebra' method, this
updated version is computed as
\begin{equation}
k_{2,10}^{15} = k_{2,10}^{16} + k_{2,15}^{16} k_{15,10}^{16} / k_{15,15}^{16}
\end{equation}
where $k_{15,15}^{16}$ is the diagonal coefficient from equation $15$. Such
a modification of the rate-coefficient has a simple physical interpretation:
to the existing set of routes transferring population from level $2$ to
level $10$, we are adding a route which takes population between these
two levels via level $15$. {\sc tracer} takes an almost completely
eliminated matrix (a $2 \times 2$ or $3 \times 3$, say) and expands the
rate-coefficients, at this very simple stage, back through the entire
elimination process to the original values that formed the matrix prior to
$N+1$ eliminations, where $N$ is the number of energy levels in the model.
By discarding the less popular routes, we can select an important subset that
form the pump of a maser, for example.

\subsection{The $1612$ MHz Pump in Zone 2}

In an energy-ordered sequence of hyperfine-resolved
energy levels in the OH molecule, the
$1612$\,MHz transition is between level $3$ and level $2$. It therefore makes
sense to start our analysis with a $3\times 3$ matrix. Solving this reduced
matrix for the populations of level $3$ and level $2$, and then forming
the inversion per magnetic sublevel, $\Delta \rho_{32}$, we obtain
\begin{equation}
\Delta \rho_{32} \! = \! \frac{{\cal N} k_{1,2}^{4}}{D} \! \left\{ \!
   \left( \!
      \frac{k_{2,3}^{4}}{3} - \frac{k_{1,3}^{4} k_{3,2}^{4}}{5 k_{1,2}^{4}}
\! \right) +
   \left( \!
      \frac{k_{1,3}^{4} k_{2,2}^{4}}{3 k_{1,2}^{4}} - \frac{k_{3,3}^{4}}{5}
\! \right)
                                               \!   \right\}
\end{equation}
where ${\cal N}$ is the total OH number density in the relevant slab, and $D$ is
a denominator, also composed of all-process rate-coefficients. The actual
coefficients for the $3 \times 3$ matrix for the time of shell detachment
and slab $46$ (in the main inverted layer, zone~2) are
\begin{equation}
                       \left[ \!
   \begin{array}{rrr}
      k_{3,3}^{4}=1.81(-4)  &  
      k_{2,3}^{4}=7.11(-5)  &  
      k_{1,3}^{4}=6.71(-5)      \\
      k_{3,2}^{4}=1.14(-4)  &   
      k_{2,2}^{4}=5.56(-4)  &
      k_{1,2}^{4}=7.73(-4)     \\
      k_{3,1}^{4}=6.69(-5)            & 
      k_{2,1}^{4}=4.85(-4)            & 
      k_{1,1}^{4}=1.35(-2)   
   \end{array}
                     \! \right]
\end{equation}
where the units are s$^{-1}$ for all coefficients.
Noting
that $k_{1,2}^{4}$, $D$ and ${\cal N}$ are all positive quantities, it is
easy to check, by substituting the values in eq.(9) into eq.(8), that the
result is a positive number, representing a $1612$\,MHz inversion.

It is possible to re-cast eq.(8) into a form more useful for studying a
maser pump across levels $2$ and $3$. The improved equation is formed by
expanding the two diagonal coefficients which appear explicitly. For
$k_{2,2}^{4}$, which represents the sum of all rate-coefficients that
take population out of level $2$, we write
\begin{equation}
k_{2,2}^{4} = k_{2,1}^{4} + k_{2,3}^{4}.
\end{equation}
Using eq.(10) and a similar expansion
for $k_{3,3}^{4}$, and defining 
$\eta = k_{1,3}^{4}/k_{1,2}^{4}$, eq.(8) is recast as
\begin{equation}
\Delta \rho_{32} = \frac{{\cal N} k_{1,2}^{4}}{D} \left\{ \!
  (1+\eta ) \! \left(
    \frac{k_{2,3}^{4}}{3} - \frac{k_{3,2}^{4}}{5}
               \right) +
    \frac{\eta k_{2,1}^{4}}{3} - \frac{k_{3,1}^{4}}{5}
                                              \!  \right\}
\end{equation}
Using eq.(11), we can see that the $1612$\,MHz pump falls into two distinct
parts. Firstly there is the `direct' contribution to the pump, represented
by the term multiplied by $(1+\eta )$, and secondly there is an `indirect'
contribution formed from the remaining terms. The indirect contribution
results from population transfers between levels $3$ and $2$ via level $1$.

\subsection{Trace for the direct pump in Zone 2}

Having established the contributions to the pump, it can be expanded by
{\sc tracer} until all the significant pump routes have been traced back
to unmodified coefficients. Of course there is a rule of diminishing returns
as one attempts to recover increasingly complete understanding of the
population transfer network, and only the strongest routes are considered
here. For the direct pump which, at the time of shell detachment, is
responsible for $58.6$ per cent of the $1612$\,MHz inversion,
we expand $k_{2,3}^{4}$, which is dominated by a single route via level $4$:
\begin{equation}
\frac{k_{2,3}^{4}}{3} - \frac{k_{3,2}^{4}}{5} \sim 
\frac{k_{2,4}^{5} k_{4,3}^{5}}{3 k_{4,4}^{5}} -
\frac{k_{4,2}^{5} k_{3,4}^{5}}{5 k_{4,4}^{5}}
\end{equation}
where this route (route $1$) is responsible for $78.6$ per cent of the
population inversion supported by the direct pump.
The remaining $21.4$ per cent of the direct pump
is carried by less important routes.
At the next stage of expansion, route $1$ breaks down
into three major contributions plus some minor routes that operate through
the $F1$ stack of levels. However, one of the major routes, operating via
level $10$ makes a small negative (anti-inverting)
contribution, and will not be considered
further here. However, this route is of some importance when discussing
the absorbing layer (see Section~5.5) so it will be labelled route $1C$
and discussed there.
The first of the inverting major routes, which we label $1A$, when
fully traced back to unmodified coefficients, yields an inversion (excluding
the multiplier $({\cal N} k_{1,2}^{4} / {\cal D})(1 + \eta )$) of
\begin{equation}
1A_{net} = \frac{k_{2,4} k_{4,7} k_{7,3}}{3 k_{7,7}^{8} k_{4,4}^{5}} -
           \frac{k_{3,7} k_{7,4} k_{4,2}}{5 k_{7,7}^{8} k_{4,4}^{5}}
\end{equation}
Route $1A$ therefore relies on a collisionally dominated transition within
a lambda-doublet via the coefficients $k_{2,4}$ and $k_{4,2}$, as
well as on a radiatively dominated transition of FIR wavelength within
the $F1$ set of energy levels. Route $1A$
contributes $39.8$ per cent of the net inversion produced
by route $1$. The strongest inverting route, which we label $1B$, operates
via level $12$, and produces $60.2$ per cent of the route $1$ inversion. This
inversion, excluding the same multiplier as for $1A$, in fully-traced form
is
\begin{equation}
1B_{net} \! = \!
\frac{k_{2,16} k_{16,12} k_{12,4} k_{4,7} k_{7,3}}{3 k_{16,16}^{17}
                 k_{12,12}^{13} k_{7,7}^{8} k_{4,4}^{5}} -
           \frac{k_{3,7} k_{7,4} k_{4,12} k_{12,16} k_{16,2}}{5 k_{4,4}^{5} 
                 k_{7,7}^{8} k_{12,12}^{13} k_{16,16}^{17}} 
\end{equation}
so we can see that route $1B$ proceeds entirely by transitions between
rotational states of OH, and we expect population
transfer in these transitions to be dominated by far-infrared radiation.

\begin{figure*}
\vspace{0.3cm}
\psfig{file=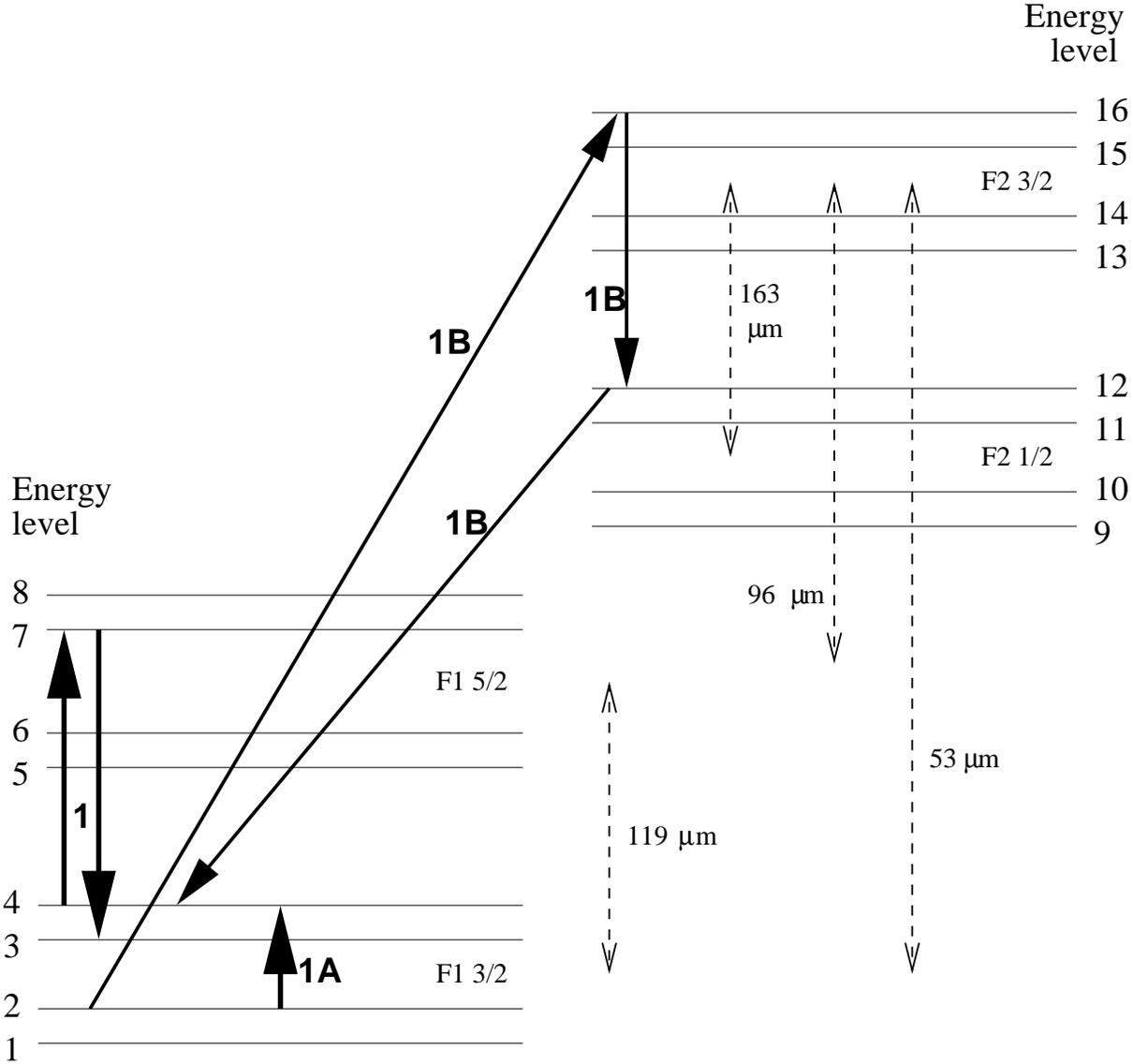,angle=0,width=16cm}
\caption{
The most important `direct' pumping routes for the $1612$\,MHz transition for
the model with a mass-loss rate of $10^{-5}$\,M$_{\odot}$\,yr$^{-1}$ at the
time of shell-detachment, and for a slab within the main inverted layer (Zone~2).
Forward routes - that is those pumping from level $2$ towards level $3$ - only are shown for simplicity (solid arrows). The most important route is the
loop marked $1B$, completed by the pair of transitions marked $1$. This is
approximately $1.5$ times as important as the route marked $1A$, which is
also completed by the pair, $1$. For further details concerning
the relative importance of particular routes, see the text of
Section~5.3. Note that the intervals between energy levels are not drawn
to scale. The $F1$ stack of levels is on the left; the $F2$ stack, on the
right. Lambda-doublets are described in Hund's case b notation. Approximate
wavelengths of far-infrared transitions are also shown (dashed arrows).
}
\label{f:figfour}
\end{figure*}

The dominant parts of the direct pump are fortunately fairly simple
because there is only one major route, route $1$, and little
branching within route $1$. All parts of
route $1$ are completed by flow via level $7$,
that is via the expansion 
$k_{4,3}^{5} \sim k_{4,7} k_{7,3} / k_{7,7}^{8}$ which contains 89.6 per cent
of $k_{4,3}^{5}$, and a similar expansion for $k_{3,4}^{5}$.
With all the coefficients in route $1$ now traced back to
their unmodified forms, it is possible to plot the pump routes on a
diagram of the OH energy-level structure. This is shown in Fig.~\ref{f:figfour}.

\subsection{Trace for the indirect pump in Zone 2}

The `indirect' pump is represented by the other term in eq.(11): 
$\eta k_{2,1}^{4} /3 - k_{3,1}^{4} / 5$. With a little re-arrangement, this
can be written in a form which explicitly shows that it represents transfer
between levels $2$ and $3$ via level $1$. The statistical weights are now
attached to the coefficients representing transfer between pairs of
levels of unequal degeneracy:
\begin{equation}
\eta k_{2,1}^{4} /3 - k_{3,1}^{4} / 5 = \frac{1}{k_{1,2}^{4}}
\left[
 \left( \frac{k_{2,1}^{4}}{3} \right) k_{1,3}^{4} -
 k_{3,1}^{4} \left( \frac{k_{1,2}^{4}}{5} \right)
\right]
\end{equation}
We begin by tracing $k_{1,2}^{4}$, and here we find simplicity:
there is a single dominant ($90.9$ per cent)
route, traceable back to original coefficients.
This route is represented by $k_{1,5} k_{5,2}/k_{5,5}^{6}$. 
We also find that $k_{2,1}^{4}$ is 
dominated by the reverse of this route. Excluding the
external multiplier, ${\cal N} / D$, the inversion
produced by the indirect pump can therefore be written as
$\epsilon_{up} k_{1,3}^{4} - \epsilon_{down} k_{3,1}^{4}$, where
$\epsilon_{up} = k_{2,5} k_{5,1} /(3 k_{1,2}^{4} k_{5,5}^{4})$ and
$\epsilon_{down} = k_{1,5} k_{5,2} /(5 k_{1,2}^{4} k_{5,5}^{4})$.

\begin{figure*}
\vspace{0.3cm}
\psfig{file=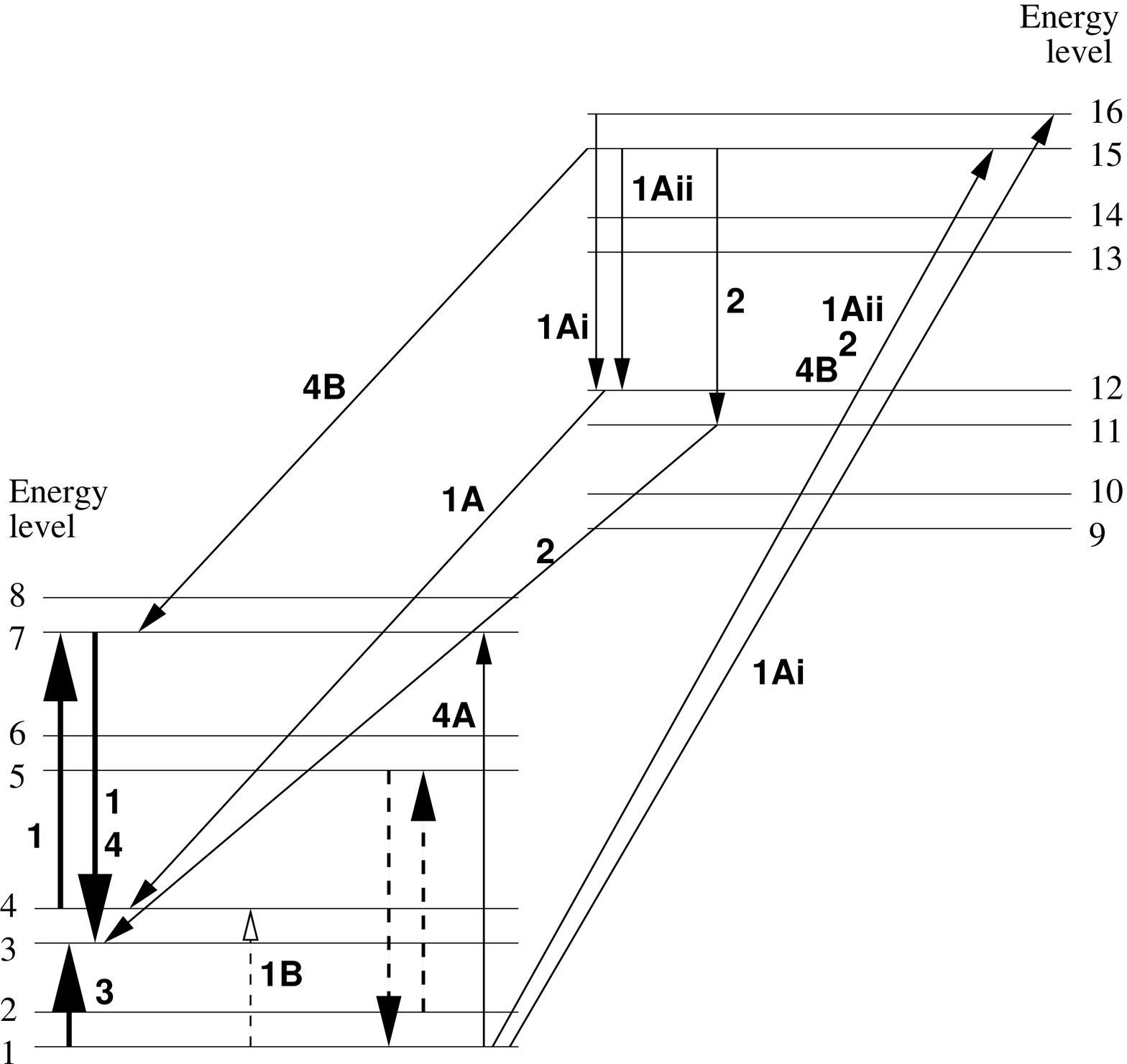,angle=0,width=16cm}
\caption{
The most important `indirect' pumping routes for the $1612$\,MHz transition for
the model with a mass-loss rate of $10^{-5}$\,M$_{\odot}$\,yr$^{-1}$ at the
time of shell-detachment, and for a slab within the main inverted layer (Zone~2). Forward routes only are shown for clarity. All routes rely on the
rapid transfer of population from level $2$ to level $1$ via level $5$ (heavy
dashed arrows). Transfer from level $1$ to level $3$ is dominated by four
routes. If routes $1$ and $2$, which are approximately equally important, are
both allocated a strength of $1.0$, then the relative importance of route $3$
is $0.81$, and route $4$ is worth 0.42. Within route $1$, $1A$ is three times
stronger than $1B$, and route $4A$ is $2.3$ times stronger than route $4B$.
For more detail about
the relative importance of particular routes, see the text of
Section~5.4. Note that the intervals between energy levels are not drawn
to scale. The $F1$ stack of levels is on the left; the $F2$ stack, on the
right. For rotational state designations and far-infrared transition
wavelengths, see Figure~\ref{f:figfour}.}
\label{f:figfive}
\end{figure*}

The remaining coefficients in the indirect system, $k_{1,3}^{4}$
and $k_{3,1}^{4}$ are much more
complicated. At the time of shell detachment, there are four 
positive contributions to
the inversion. The two most important contributions, each with a value
of $2.53 \times 10^{-7}$\,s$^{-1}$, are pumps via level $4$ (route $1$) and level
$11$ (route $2$) . There is also an important route using the unmodified coefficients,
$k_{1,3}$ and $k_{3,1}$ (route $3$). The group $\epsilon_{up} k_{1,3} -
\epsilon_{down} k_{3,1}$, corresponding to this contribution has the
value $2.06 \times 10^{-7}$\,s$^{-1}$. Also significant is a route via
level $7$ (route $4$), which contributes $1.07 \times 10^{-7}$\,s$^{-1}$. Two additional
contributions via levels $5$ and $9$ also carry significant amounts of
population, but make negative contributions to the inversion, and will not
be traced further. The strong route via level $4$, which we have called route
$1$ of the indirect system, can be further traced back to a dominant
route via level $12$ (route $1A$) and a route which uses the unmodified
coefficients $k_{1,4}$ and $k_{4,1}$ (route $1B$). Route $1A$
generates $75.3$ per cent of the inversion produced by
route $1$; the remainder is pumped by route $1B$. Route $1A$ can be further
broken down into contributions via level $16$ (route $1Ai$) and, via level
$15$ (route $1Aii$). Route $1Ai$ is $3.41$ times more potent than route
$1B$. Route $2$ can be traced back from level $11$ to include level $15$,
but does not involve further branching, whilst route $4$ branches into
two components, one going no higher than level $7$ in the $F1$ stack of
levels and a second branch operating via level $15$. The latter branch
(route $4B$) carries $30$ per cent of route $4$. With these traces
complete, the subset of the most important indirect pumping routes
is shown in Fig.~\ref{f:figfive}. 

\subsection{The absorbing layer}

The next region we consider is slab number $20$, which is within the 
absorbing layer. Here, the exact coefficients, the analogues of those in
eq.(9), are,
\begin{equation}
                       \left[ \! \!
   \begin{array}{rrr}
      k_{3,3}^{4}=9.03(-5)  &  
      k_{2,3}^{4}=2.83(-5)   &  
      k_{1,3}^{4}=3.62(-5)      \\
      k_{3,2}^{4}=5.26(-5)  &   
      k_{2,2}^{4}=1.17(-4)  &
      k_{1,2}^{4}=1.66(-4)     \\
      k_{1,1}^{4}=3.77(-5)               & 
      k_{2,1}^{4}=8.92(-5)               & 
      k_{3,1}^{4}=2.63(-3)   
   \end{array}
                   \! \! \right]
\end{equation}
and comparison with eq.(9) immediately shows that a major change is a
diminution of all the coefficients, indicating a
generally less efficient transfer of population.

 When the investigation
with {\sc tracer} is begun, we find that both the direct and indirect parts
of the pump are now anti-inverting: the expression for the direct
pump reduces to $-1.223 \times 10^{-6}$\,s$^{-1}$, compared to
$+1.081 \times 10^{-6}$\,s$^{-1}$ for the example in Zone~2; the indirect
pump yields $-1.019 \times 10^{-6}$\,s$^{-1}$ in the absorbing layer,
instead of $+6.457 \times 10^{-7}$\,s$^{-1}$ in the chosen slab from
Zone~2. Overall, we find a similar pattern of routes to
that available in the inverted case, but with significant changes. Route
$1B$ in Fig.~\ref{f:figfour} has become generally weaker, but remains an inverting route.
A partial explanation for the change to absorption is that route $1A$ has
gone from being strongly inverting in Zone~2 to strongly anti-inverting in
the absorbing layer. As the rate-coefficients $k_{2,4}$ and $k_{4,2}$ are
almost equal (as expected for a collision-dominated transtion with a small
energy gap) the main change to the coefficients which appear in eq.(13)
must involve the coefficients which link levels $3$ and $4$ to level $7$.
On examining these coefficients in detail, the switch
to inversion in this case results from the $3 \rightarrow 7$ transition
becoming relatively optically thinner than the $4 \rightarrow 7$ 
transition, in moving
from Zone~2 to the absorbing layer.

The second route that produces strong anti-inversion was briefly introduced
in Section~5.3 as route $1C$, noting that it made an anti-inverting
contribution, even in Zone~2. In the absorbing layer this route, given
by
\begin{equation}
1C_{net} \! = \!
\frac{k_{2,10} k_{10,14} k_{14,4} k_{4,7} k_{7,3}}{3 k_{14,14}^{15}
                 k_{10,10}^{11} k_{7,7}^{8} k_{4,4}^{5}} -
           \frac{k_{3,7} k_{7,4} k_{4,14} k_{14,10} k_{10,2}}{5 k_{4,4}^{5}      k_{7,7}^{8} k_{10,10}^{11}
                 k_{14,14}^{15}}
\end{equation}
has values of $3.12\times 10^{-7}$\,s$^{-1}$ for the first (forward) part, and  $6.99\times 10^{-7}$\,s$^{-1}$ for the second
(reverse) part. In Zone~2, the analogous values are $3.70\times 10^{-6}$\,s$^{-1}$ and $3.84\times 10^{-6}$\,s$^{-1}$. Part of
the change towards increased net absorption is explained by the route
between levels $3$ and $4$ via level $7$, which route $1C$ shares with
$1A$, but there is also a contribution from the group
$k_{2,10} k_{10,14} k_{14,4}/(3 k_{10,10}^{11} k_{14,14}^{15})$ and its
reciprocal, which has also become more anti-inverting. This is definitely
not due to the $53$\,$\mu$m transitions, $2\rightarrow 10$ and $10\rightarrow 14$,
but results from optical depth changes in the $10\rightarrow 14$ transition
at $163$\,$\mu$m. 

Changes to indirect pumping routes are also responsible for about half the
anti-inverting power in the absorbing layer. Route $1A$ 
in Fig.~\ref{f:figfive} remains
inverting, but with reduced effectiveness, because it includes the link
from level $4$ to level $3$ via level $7$. This same link causes route
$1B$ to fall into an anti-inverting state. For route $3$ the link from
level $2$ to level $1$ via level $5$ plays a similar role to the
$4 \rightarrow 7 \rightarrow 3$ sequence. Both operate at a wavelength
of $119$\,$\mu$m, and differential changes in the optical depth of the
upward and downward parts push the whole scheme towards net absorption. The
rest of the picture for the indirect part of the pump is much more
complicated: the absorption zone has no equivalents of routes $2$ or
$4$ with significant strength, but these have been replaced by a complicated
web of anti-inverting routes. Of particular importance is a route from
level $2$ to level $10$, and thence to level $3$ via level $1$, in which
the group
\begin{equation}
Q5 = \frac{k_{2,10} k_{10,1} k_{1,3}}{3 k_{1,2}^{4}
                 k_{10,10}^{11}} -
           \frac{k_{3,1} k_{1,10} k_{10,2}}{5 k_{10,10}^{11}
                 k_{1,2}^{4}}
\end{equation}
has the value $-1.69\times 10^{-7}$\,s$^{-1}$.

\subsection{the outer emission layer, zone~1}

The main qualitative differences between this zone and the inner inverted
zone, zone~2, lie in the relative importance of the direct and indirect
contributions to the inversion, and in the overall complexity. In zone~1, the
indirect pump contributes 60.2 per cent of the inversion. The direct pump
is similar to that shown in Fig.~\ref{f:figfour}, but has increased complexity due to
additional significant routes. In particular, 
route $1A$ now provides some $80$\% of 
the inversion generated by route $1$, and there is a weaker companion 
to route $1B$, which operates via level $15$, rather than $16$, The direct
route from level $4$ to level $12$, and its reciprocal, is also
significant, whilst, importantly, the anti-inverting route $1C$ has diminished.

In the indirect route, we have a similar concentration towards the route
which uses only $119$\,$\mu$m radiation: the route from level $2$ to level $3$
via level $5$ (route 3 in Fig.~\ref{f:figfive}) provides $94$\% of the indirect inversion. 

\subsection{zone~2 forty years after detachment}

Here, we consider the changes to the pumping scheme which result from $40$\,yr
of expansion of the shell, counting from `detachment', where significant
mass loss ends. We consider a layer at the same absolute depth as in
Section~5.3 and 5.4.  
The first observation is that, at this advanced time, the
direct pump $(1+\eta )(k_{2,3}^{4} - k_{3,2}^{4})$ is actually mildly
larger, in absolute terms, than at shell detachment. The significant reduction in maser output, visible in Fig.~\ref{f:figone}, must therefore be due to the loss of
effectiveness in the indirect pump. This view is confirmed on examination
of the coefficients for the indirect pump, which now has a very small
negative value. On a more
detailed examination of the pump routes, there is also an obvious reason
for the change to the indirect pump: both the strongest pump routes, route $1$
and route $2$ in Fig.~\ref{f:figfive} are now strongly anti-inverting. By comparison, the
changes to routes $3$ and $4$ are small. For route $1$, the value of the
group $\epsilon_{up} k_{1,4}^{5} k_{4,3}^{5} -
       \epsilon_{down} k_{3,4}^{5} k_{4,1}^{5}$ has fallen from
$2.53 \times 10^{-7}$\,s$^{-1}$ at shell detechment to the strongly
anti-inverting value of $-6.62 \times 10^{-7}$\,s$^{-1}$ forty years later.
For route $2$, the value of the expression
$\epsilon_{up} k_{1,11}^{12} k_{11,3}^{12} -
       \epsilon_{down} k_{3,11}^{12} k_{11,1}^{12}$ has fallen, over the
same period of time, from $2.53 \times 10^{-7}$\,s$^{-1}$ to 
$-8.52 \times 10^{-7}$\,s$^{-1}$. Although a route via level $9$ has become
inverting, it is not nearly enough to compensate for the collapse of
routes $1$ and $2$.

Within route $1$, there are the $A$ and $B$ parts. The $B$-part of the route,
which involves the lambda-doublet coefficients $k_{1,4}$ and $k_{4,1}$ still
has $61$ per cent of the inverting effect it had at shell detachment. The
big changes are in the $1A$ routes: the route $1Ai$, via level $16$ retains
only $37$ per cent of its inverting effect, whilst route $1Aii$, via
level $15$, has become strongly anti-inverting. We note that all three
of the routes that have experienced major decreases in their inverting
power require $53$\,$\mu$m radiation for the initial stage of the
forward (inverting) part of the route.

To test the hypothesis that the initial absorption of $53$\,$\mu$m 
radiation is
the process most affected by the time evolution of the envelope, we can look
at the changes to the upward and downward parts of each group separately.
For route $1B$, we find that the downward part of the route, represented
by the group $\epsilon_{down} k_{3,11} k_{11,15} k_{15,1} / (k_{11,11}^{12}
k_{15,15}^{16})$ has increased from $3.51 \times 10^{-7}$\,s$^{-1}$ to
$6.40 \times 10^{-7}$\,s$^{-1}$, whilst the upward counterpart has
fallen to just $7.28 \times 10^{-8}$\,s$^{-1}$. The fall in the upward
part of the route therefore has the stronger effect. For route $2$, both the
upward and downward parts of the routes reduce with time, but the upward
part reduces more: the upward group changes from $1.37\times 10^{-6}$\,s$^{-1}$
to $1.76 \times 10^{-7}$\,s$^{-1}$, whilst the value of the
downward expression changes from $1.13 \times 10^{-6}$\,s$^{-1}$ to
$8.00 \times 10^{-7}$\,s$^{-1}$. 

Of the three coefficients in the upward group, only $k_{1,15}$ is common to
route $1Aii$ and route $2$, and it is also the only one which shows a
strong change over the $40$\,yr of expansion. This coefficient itself falls
from $1.993 \times 10^{-5}$\,s$^{-1}$ at shell detachment to
$3.906 \times 10^{-6}$\,s$^{-1}$ at the later time. The value of its
downward counterpart, $k_{15,1}$, is not noticeably different from its
A-value at the later time, indicating that this transition has a negligible
collisional component, and that it has become optically very thin. Therefore,
although other transitions have significant effects, the largest contribution
to the decay of maser radiation is the decreasing mean intensity in
the radiation field at
$53$\,$\mu$m.

\begin{figure*}
\vspace{0.3cm}
\psfig{file=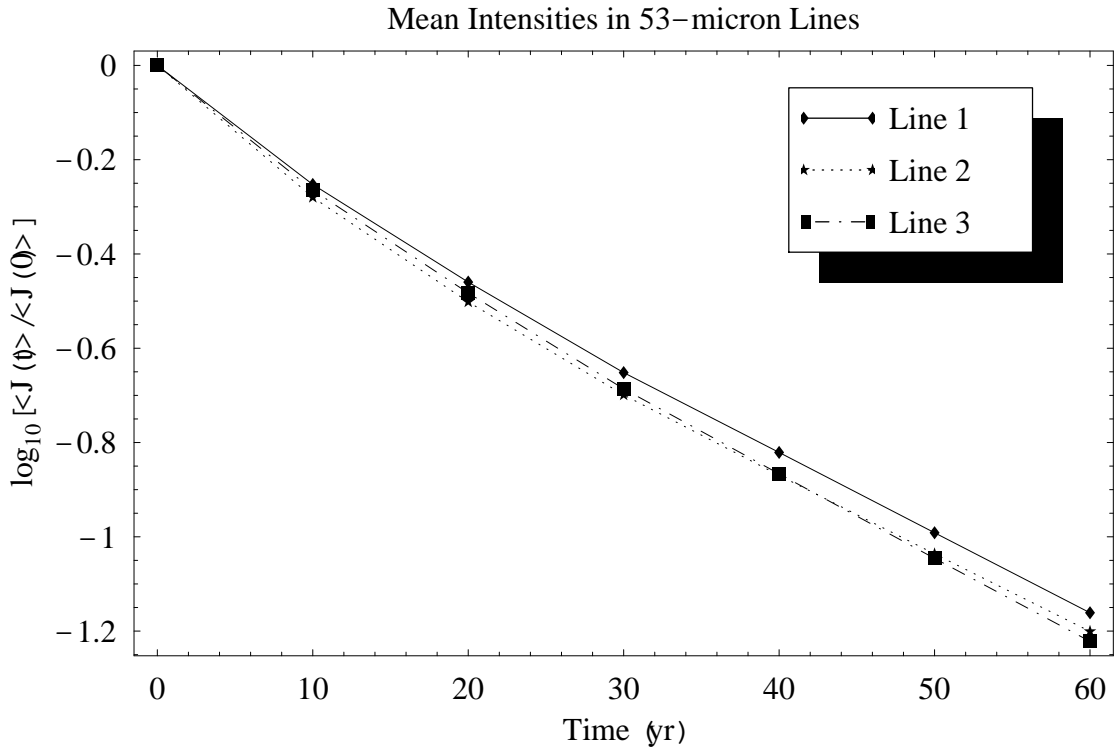,angle=0}
\caption{
Decay of the angle-averaged intensity, at line centre, of the three important
pumping lines at $53$\,$\mu$m. These lines are, line $1$ (levels 1-16 in
Fig.~\ref{f:figfive}; line $2$ (levels 2-16 in Fig.~\ref{f:figfour};
line $3$ (levels 1-15 in Fig.~\ref{f:figfive}). Absolute values of
$<J(0)>$ are, respectively, $2.601 \times 10^{-12}$, $1.446 \times 10^{-12}$
and $2.598 \times 10^{-12}$\,W\,m$^{\rm -2}$\,Hz$^{\rm -1}$. The time origin
is at shell detachment.
}
\label{f:figsix}
\end{figure*}

To test the hypothesis that a lack of $53$\,$\mu$m radiation is responsible for
the loss of the $1612$\,MHz pump, we plot in Fig.~\ref{f:figsix} the
mean (that is angle, but not frequency, averaged) intensities, at
line centre, of the three
$53$\,$\mu$m lines which play important roles in Fig.~\ref{f:figfour} and
Fig.\ref{f:figfive}. These plots are for the same mass-loss rate 
($10^{-5}$\,M$_{\odot}$\,yr$^{-1}$) as used for the {\sc tracer} analysis,
and for the same depth in zone~2 ($1.289 \times 10^{14}$\,cm). The mean
intensity has been chosen as the function to plot because it is the
variable on which the radiative part of the pump directly depends via its
products with the Einstein B-coefficients. The three chosen lines are 
line $1$ ($53.2532$\,$\mu$m; levels 1-16 in Fig.~\ref{f:figfive}), line $2$
($53.2537$\,$\mu$m; levels 2-16 in Fig.~\ref{f:figfour}) and line $3$
($53.2539$\,$\mu$m; levels 1-15 in Fig.~\ref{f:figfive}). In the radiation
transfer code, lines $2$ and $3$ form an overlapping pair, and
all three lines were found in absorption. Lines $1$ and $3$
are believed responsible for the collapse of the strongest part of the
pumping scheme in Fig.\ref{f:figfive}. From Fig.~\ref{f:figsix}, we observe
that the mean intensity in all three lines, relative to the value at shell
detachment, follows a very similar quasi-exponential decay. This tends to
confirm that falling energy density in these lines is responsible for the
decay of the `indirect' part of the $1612$\,MHz pump.

\section{Discussion}

We have shown that a model comprising a hydrodynamic 
and photochemical model of the shell of
an OH/IR star, and a radiative transfer code for the continuum and OH lines,
can reproduce the decay of $1612$\,MHz maser emission observed in OH/IR stars. The most pertinent results from the model come from
mass loss rates between $1.0\times 10^{-6}$\,M$_{\odot}$\,yr$^{-1}$ and
$2.0\times 10^{-5}$\,M$_{\odot}$\,yr$^{-1}$, where the decay time of the
masers varies almost linearly with the
mass-loss rate. Decay half-lives derived from unsaturated inversions are
comparable with, but slightly larger than, those found observationally.
However, if saturation is taken into account, some time would be taken
up reducing the degree of saturation, with little change in maser intensity,
and thus compressing the observable decay into somewhat
shorter intervals than those found with the present model.

The inversions that provide the maser emission are supported in a series
of layers within the circumstellar envelope. There are four basic zones:
a physically, and optically, thin outer emitting layer which provides only
a small amount of the maser gain; an absorbing zone which is sandwiched
between the outer emitting zone and a deeper inverted layer, and finally
a deep absorbing layer which runs inwards to the edge of the modelled
part of the shell. As the shell evolves, following detachment, the zones
all tend to move deeper into the shell, and away from the surface (the
closest point to the observer). Detailed tracing of the population flow
through the OH energy levels has been carried out for representative
points in the two inverted layers, and the intervening absorbing layer, with
the aim of determining what processes are ultimately responsible for the
maser decline.

We find that the strongest pumping routes use FIR transitions at $53$\,$\mu$m
to cross from the $F1, J=3/2$ ground state to $F2, J=3/2$.
This is followed by
decay to $F2, J=1/2$, before a `cross-stack' return to the ground state.
This type of pump is essentially a modification of the scheme proposed
by Elitzur, Goldreich \& Scoville \shortcite{egs76} (EGS): the initial stage
of the pump relies on $53$\,$\mu$m radiation, rather than the 
$35$\,$\mu$m transition used in EGS. However, the variant with a
$53$\,$\mu$m initial stage is suggested in Elitzur \shortcite{el81} as
an alternative which is likely to 
operate in cooler envelopes \cite{dic87}. We note that
in the model used in the present work, the dust temperature in Zone~2
is typically $25$-$40$\,K, rather than the $100$-$200$\,K needed to drive the
$35$\,$\mu$m version. Routes that use $35$\,$\mu$m radiation are traced
in our model, but were never found to be strong enough to include in the
discussion.
In this connection, the $35$\,$\mu$m line was a
target in several observations with the ISO satellite. Apart from the
Galactic Centre, $35$\,$\mu$m lines were detected in absorption by
Neufeld et al. \shortcite{neu99} towards VY~CMa, and by Sylvester et al.
\shortcite{syl97} towards IRC+10420, providing strong support for the
EGS model. However, neither of these objects are typical of the moderate to
low-mass OH/IR stars considered by the present work. Sylvester et al.
also searched for the $53$\,$\mu$m lines, but there were no detections
either in emission or absorption. $53$\,$\mu$m radiation could, however,
contribute pumping at a level up to 50\% of the $35$\,$\mu$m route in
IRC+10420 \cite{syl97}. Szczerba, He \& Chen \shortcite{szc03} looked at 
81 OH/IR sources from the ISO archive, all with data covering the OH
$35$\,$\mu$m transitions. No additional detections were found. A possible
interpretation is that most objects in the ISO archive have envelopes
which utilize the alternative EGS scheme that
is based on $53$ micron radiation.

Spatially, the effects which divide the envelope into the various zones of
emission and absorption appear to be controlled, in general terms, by
differential optical depth effects in both $53$\,$\mu$m and
$119$\,$\mu$m lines. Moving towards the surface of the envelope, the
transition from Zone~2 to the absorbing layer appears to be controlled 
mainly by changes in the optical depth of $119$\,$\mu$m transitions, since
pumping routes powered by $53$\,$\mu$m radiation remain important, and
tend less readily to anti-inversion than those routes which use
$119$\,$\mu$m transitions only. The boundary between the absorbing layer
and Zone~1 appears to be due mainly to a reduction in the strength
 of the $53$\,$\mu$m pumping
routes. However, an additional change in the optical depths 
of $119$\,$\mu$m lines must also take place, so that these now drive
inversions in Zone~1, which is not so in the absorbing layer.

When the masers decline, we observe that the pumping routes which are lost,
that is, which become anti-inverting or significantly less inverting
over $40$\,yr, are those
which require a $53$\,$\mu$m upward transition. Population transfer in these routes is dominated
by interactions with far infrared radiation. By contrast routes involving
transitions within lambda doublets, which are collisionally dominated,
maintain their inverting strength quite well over forty years of expansion.
We therefore conclude that it is the changes in the radiation field, rather
than the kinetic temperature and density in the envelope, which are
mainly responsible for the observed fall in maser gain at $1612$\,MHz.

Several pumping routes which depend upon FIR radiation still operate well
in the envelope $40$\,yr after shell detachment. Those which become
anti-inverting seem to require the additional feature of an initial
upward transition driven by radiation at a wavelength of $53$\,$\mu$m, which
is the most energetic radiation that is important for the traced pumping
schemes. The fact that the loss of the most energetic radiation is important
suggests that the underlying change in the envelope, which is responsible
for the decay in maser gain, is a cooling of the dust-generated radiation
field. This has been confirmed by
plotting directly the mean intensity in three $53$\,$\mu$m lines as a
function of time.

\subsection*{ACKNOWLEDGMENTS}

MDG acknowledges PPARC for financial support under the UMIST astrophysics
2002-2006 rolling grant, number PPA/G/O/2001/00483. This research is supported by the National Astronomy and Ionosphere
Center, which is operated by Cornell University under a cooperative
management agreement with the National Science Foundation.
The authors would also
like to thank Katherine Lynas for tracing some of the pumping routes, 
which appear in Figures \ref{f:figfour} and \ref{f:figfive}.

\end{document}